\documentclass[twocolumn]{aastex631}

\begin{document}
\newcommand{\mycomment}[1]{}
\newcommand{\ergs}{\hbox{erg\,s$^{-1}$}}
\newcommand{\msun}{\rm M_\odot}
\newcommand{\zsun}{\rm Z_\odot}

\shorttitle{Delayed Feedback at Low Metallicity}
\shortauthors{Jecmen \& Oey}

\received{May 8, 2023}
\revised{September 19, 2023}
\accepted{October 12, 2023}
\submitjournal{AAS}

\title{Delayed Massive-Star Mechanical Feedback at Low Metallicity}

\author{Michelle C. Jecmen}
\affiliation{Department of Astronomy, University of Michigan, 1085 S. University Ave, Ann Arbor, MI 48109-1107}

\author{M. S. Oey}
\affiliation{Department of Astronomy, University of Michigan, 1085 S. University Ave, Ann Arbor, MI 48109-1107}

\begin{abstract}
The classical model of massive-star mechanical feedback is based on effects at solar metallicity $(\zsun)$, yet feedback parameters are very different at low metallicity. Metal-poor stellar winds are much weaker, and more massive supernova progenitors likely collapse directly to black holes without exploding. Thus, for $\sim0.4\ \zsun$ we find reductions in the total integrated mechanical energy and momentum of $\sim40$\% and 75\%, respectively, compared to values classically expected at solar metallicity. But in particular, these changes effectively delay the onset of mechanical feedback until ages of $\sim10$ Myr. Feedback from high-mass X-ray binaries could slightly increase mechanical luminosity between ages 5--10 Myr, but it is stochastic and unlikely to be significant on this timescale. Stellar dynamical mechanisms remove most massive stars from clusters well before 10 Myr, which would further promote this effect; this process is exacerbated by gas retention implied by weak feedback. Delayed mechanical feedback implies that radiation feedback therefore dominates at early ages, which is consistent with the observed absence of superwinds in some extreme starbursts. This scenario may lead to higher star-formation efficiencies, multiple stellar populations in clusters, and higher Lyman continuum escape. This could explain the giant star-forming complexes in metal-poor galaxies and the small sizes of OB superbubble shells relative to their inferred ages. It could also drive modest effects on galactic chemical evolution, including on oxygen abundances. Thus, delayed low-metallicity mechanical feedback may have broad implications, including for early cosmic epochs.
\end{abstract}


\section{Introduction} \label{sec:intro}
Mechanical feedback from massive stars and young star clusters plays a pivotal role in the evolution of star-forming galaxies. Supersonic stellar winds from OB stars and their supernovae (SNe) release large amounts of mechanical energy ($\sim 3 \times 10^{38}$  erg  s$^{-1}$ per solar mass of stars \citep{Keller2014} which strongly affect the surrounding interstellar medium (ISM). The shocks generate superbubbles that are pressure-driven by a hot ($>10^6$ K), low-density component which piles up cooler ISM into large, ionized and/or neutral shells.  In starbursts, the super star clusters (SSC) drive galactic superwinds that blow hot gas and newly synthesized supernova products into the circumgalactic medium (CGM). Thus, massive-star feedback can be responsible for the morphology, kinematics, ionization balance, and metallicity of the ISM and CGM. Furthermore, expanding superbubbles and superwinds can trigger second, and even third, waves of star formation as they interact with surrounding ISM \citep[e.g.,][]{Oey2005}. 

This classical model for mechanical feedback was formulated for effects at solar metallicity and is approximately constant over the cluster lifetime.  However, this is not the case at low metallicity, as the initial mechanical feedback of a low metallicity cluster is at least an order of magnitude lower than its maximum value well into the cluster lifetime \citep{Leitherer2014}.  In this work, we show that this effect is likely to be important and could fundamentally impact galaxy evolution in multiple ways.

It is well understood that low metallicity mechanical feedback is decreased for the first $\sim$3 Myr due to weak stellar winds \citep[see, e.g., the review by][]{Vink2022}. Recent observational and theoretical evidence suggest that widely used low-metallicity stellar wind mass-loss rates are too high. Mass-loss rates obtained from H$\alpha$ and UV absorption features reveal values that are smaller than theoretical mass-loss rates from, e.g., \citet{Vink2001} by a factor of $\sim$3 \citep{Bouret2012,Surlan2013,Puls2008}. Recent numerical models continue to support this moderate decrease \citep[e.g.,][]{Gormaz2022,Vink2021}. 
However, recent studies which account for wind micro-clumping and weak winds show a much more dramatic decrease from Vink's (2001) formula of up to one, or even two, orders of magnitude \citep{Bjorklund2022,Rickard2022, Ramachandran2019}. Observations in the Small Magellanic Cloud also support much weaker winds \citep{Ramachandran2019} than expected.

Another major effect at low metallicity is that SNe start later than at solar values. The classical paradigm is that all massive stars above $\sim$8 $M_\odot$ explode, starting at an age of $\sim$3 Myr and continuing steadily until the low-mass limit for core collapse is reached around 45 Myr. However, core-collapse supernova (CCSN) models find that many potential progenitors do not explode but instead form black holes (BH). This effect is well established for low metallicity \citep{Zhang2008, O'Connor2011, Patton2020}. There is no strong consensus on the maximum progenitor mass for SN explosions. \citet{Sukhbold2016} find that only 10\% of stars more massive than 20 $\msun$ successfully explode, and \citet{O'Connor2011} find a strict upper limit of 30 $M_\odot$ at very low metallicities.

Additionally, mechanical feedback models overlook dynamical processes, which remove stars, particularly massive stars, before they contribute to the total mechanical feedback of the cluster. These processes are driven by massive binaries, which are more prevalent at low metallicity. When star formation ends, a cluster is not in dynamical equilibrium. As it re-virializes, a significant number of stars will naturally evaporate \citep[e.g.,][]{Brinkmann2017}. In addition, the massive stars segregated in the dense cluster core are likely to be dynamically ejected \citep[e.g.,][]{Oh2015,Oh2016}. If ejected early enough or fast enough, the stars can travel to distances where their mechanical feedback no longer contributes to the aggregate as manifested by a superbubble or superwind. 

Taken together, these effects imply that mechanical feedback is profoundly different at low metallicity than assumed by the classical paradigm. In this paper we show that the combination of weak stellar winds, fewer supernovae, and the removal of stars by dynamical processes effectively delays the onset of mechanical feedback until cluster ages of $\sim 10$ Myr. This has profound effects on the character of massive-star feedback and implies that radiation dominates over mechanical feedback at early ages, which has significant consequences, many of which remain to be understood.

\section{Starburst99 Models} \label{sec:SB99} 

We use Starburst99, a well-established evolutionary synthesis code, to model the mechanical feedback of star-forming galaxies with varying SN progenitor masses \citep{Leitherer2014}. It was previously thought that all stars between 8 and 120 $M_\odot$ end their lives as SNe. However, it is now believed that many low-metallicity progenitors with extended core structure experience direct core collapse without SNe \citep[e.g.,][]{O'Connor2011, Zhang2008, Heger2003, Sukhbold2016}. The explodability of a progenitor depends on its stellar structure, which is a function of mass and metallicity. 

To explore the effect of limiting SNe from massive progenitors, we calculate the mechanical luminosity and momentum injection rate over time for three cluster models with varying SN progenitor masses and metallicities. In what follows, the ``classical model'' has a maximum SN progenitor mass of 120 $\msun$ and solar metallicity. The ``unrestricted SNe model'' is identical to the classical model but has subsolar metallicity. The ``restricted SNe model'' has a maximum SN progenitor mass of 20 $\msun$, in line with the predictions of \citet{Sukhbold2016}, and has subsolar metallicity. 

For all models we retain the typical minimum SN progenitor mass of 8 $\msun$, and assume that all stars with masses between the minimum and maximum values explode. We use two stellar evolutionary models published by the Geneva group: one excluding stellar rotation and one with a rotational velocity of 40\% of break-up velocity \citep{Georgy2013, Ekstr2012}. All models include stellar rotation unless stated otherwise. The models at subsolar metallicity have heavy-element abundances of $Z=0.002$ and $Z=0.004$ for their evolutionary model \citep{Georgy2013} and atmospheric model \citep{Meynet1994}, respectively. All models use instantaneous star formation with a Kroupa initial mass function for a 10$^6$ $\msun$ cluster.

\subsection{Mechanical Feedback}

\begin{figure*}
    \includegraphics[width=3.4in]{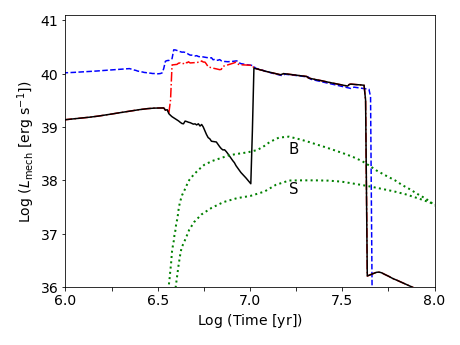} ~~~
    \includegraphics[width=3.4in]{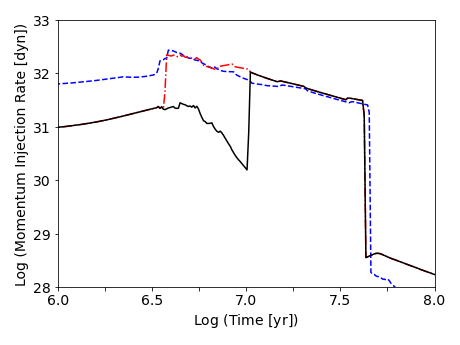}
  \caption{Starburst99 models of mechanical luminosity (left) and momentum injection rate (right) from winds and SNe for a 10$^6\ \msun$ cluster \citep[cf.][]{Leitherer2014}. The classical (blue) and subsolar unrestricted SNe (red) models have SN progenitor masses 8 -- 120 $\msun$. The subsolar, restricted SNe (black) model has SN progenitor masses of 8 -- 20\ $\msun$. The stellar evolutionary tracks include rotation for both solar \citep{Ekstr2012} and subsolar \citep{Georgy2013} metallicity. The green dotted lines in the left panel show mechanical feedback for the binary population synthesis models of \citet{Rappaport2005} for an HMXB population of $10\ \msun$ black holes with secondary stars drawn from a \citet{Salpeter1955} IMF. Models S and B show $L_{\rm mech} = L_{\rm Edd}$ and $10 L_{\rm Edd}$, respectively (see text).}
  \label{fig:Feedback}
\end{figure*}

Figure \ref{fig:Feedback} shows the mechanical luminosity (left) and momentum injection rate (right) over time due to stellar winds and SNe for different limits on explodability and metallicity. The dashed blue line shows the classical mechanical feedback scenario where strong stellar winds dominate until SNe start at $\sim$3 Myr. The magnitude of the stellar wind contribution roughly equals that of the SNe contribution, thereby generating a fairly constant value with time. This is not the case for the subsolar models. The solid red and black lines show the subsolar restricted and unrestricted SNe models, respectively. Since stellar winds are significantly weaker at low metallicity, the start of SNe is clearly identifiable by a sharp increase of an order of magnitude or more in both plots. Since the restricted SNe model limits the SN progenitor range to masses $< 20\ \msun$, the onset of SNe is delayed to an age of $\sim$10 Myr. Stellar winds now dominate the mechanical luminosity and the momentum injection rate for this much longer initial period. This causes a great reduction in both the mechanical luminosity and momentum injection rate because, not only are low metallicity stellar winds less powerful than SNe, but also after $\sim$3 Myr, as massive stars begin to expire, the remaining lower-mass stars have exponentially weaker stellar winds. Furthermore, we expect the mechanical feedback from stellar winds to be even lower than estimated here, since the adopted mass-loss rates are now thought to be too high \citep[e.g.,][]{Vink2022,Bjorklund2022,Gormaz2022,Rickard2022,Ramachandran2019}. 

\begin{figure*}
    \includegraphics[width=3.4in]{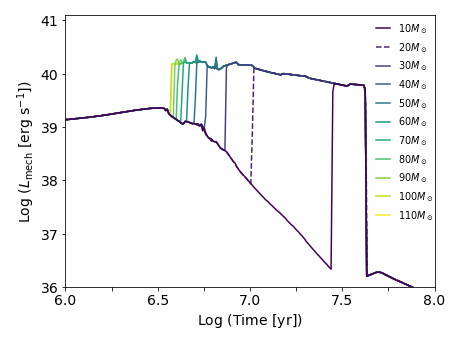} ~~~
    \includegraphics[width=3.4in]{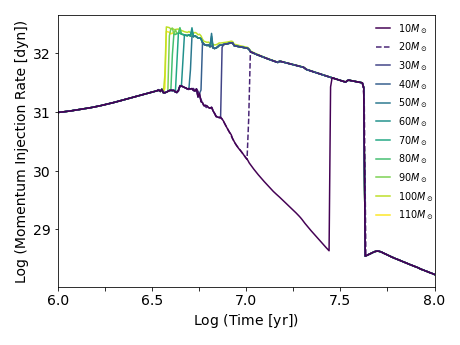}
  \caption{Mechanical luminosity (left) and momentum injection rate (right) over time for a subsolar 10$^6\ \msun$ cluster. Models with varying maximum SN progenitor masses are shown with different colors as indicated, in increments of 10 $\msun$. The model with a maximum SN progenitor of 20 $\msun$ (dashed line) corresponds to the default subsolar, restricted SNe model in Figure \ref{fig:Feedback}.}
  \label{fig:BHCutoffs}
\end{figure*}

Thus, these combined effects delay the start of strong, SN mechanical feedback until a cluster age of $\sim 10$ Myr.
The effect of varying the maximum SN progenitor mass is shown in Figure \ref{fig:BHCutoffs} for mechanical luminosity and momentum injection rate, for subsolar models.  The models with less massive maximum SN progenitors have exponentially larger delays in the onset of SN feedback, and explodability is also stochastic. Thus, we should take 10 Myr as a nominal value for the feedback delay. 

\begin{figure*}
    \includegraphics[width=3.4in]{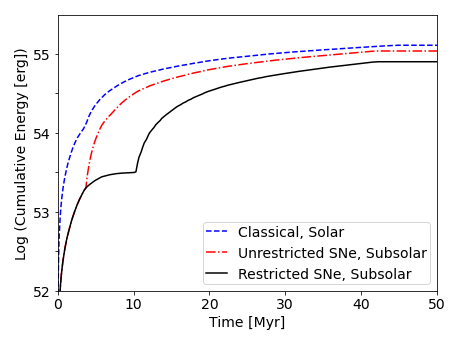} ~~~
    \includegraphics[width=3.5in]{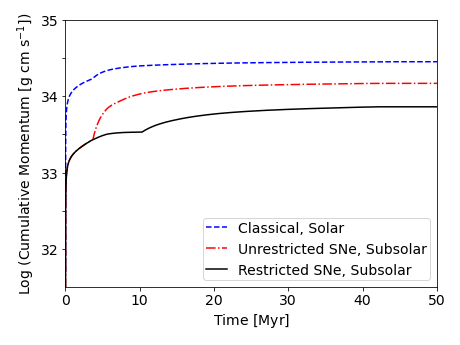}
  \caption{Total cumulative integrated energy (left) and total momentum injected (right) from winds and SNe. Models are the same as in Figure \ref{fig:Feedback}.}
  \label{fig:Cumulative}
\end{figure*}

Figure \ref{fig:Cumulative} shows the total integrated energy (left) and cumulative momentum (right) for the models in Figure \ref{fig:Feedback}. The effect from weak, low metallicity stellar winds is shown by the difference between the classical model (blue) and the subsolar, unrestricted SNe model (red). The effect from only restricting SN progenitor masses is shown by the difference between the subsolar, unrestricted SNe model (red) and the restricted SNe model (black). These differences are apparent in the total integrated energy of the cluster at early ages and become less differentiated as the cluster ages. However, for the total momentum injected, the difference between the models remains large over the duration of mechanical feedback input, since, as seen in Figure~\ref{fig:Feedback}, the momentum injection rate decreases significantly, and more steeply than, the energy injection rate. Thus, the initial momentum loss due to the absence of early SNe causes a significant decrease in the total, cumulative momentum feedback that the system never recovers.

\begin{figure*}
    \includegraphics[width=3.4in]{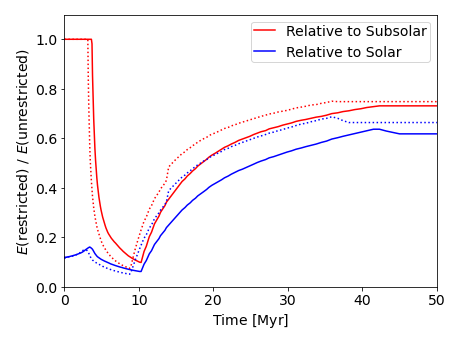} ~~~
    \includegraphics[width=3.4in]{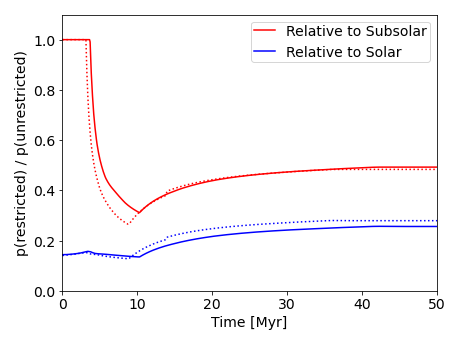}
  \caption{The total cumulative integrated energy (left) and total momentum injected (right) for the subsolar, restricted SNe model normalized by the classical, solar model (blue) and unrestricted, subsolar model (red). The solid and dotted lines show models including and excluding stellar rotation, respectively.}
  \label{fig:Normalized}
\end{figure*}

Figure \ref{fig:Normalized} gives the cumulative energy ($E$) and cumulative momentum ($p$) for the restricted SNe model at subsolar metallicity, normalized by the classical model (blue) and subsolar unrestricted SNe model (red). For the first $\sim$10 Myr, both the cumulative energy and momentum of the restricted SNe model are only $\sim10$\% that of the classical model. Ultimately, the cumulative energy of the restricted SNe model reaches $\sim$ 60\% of the classically expected energy. In comparison, the cumulative momentum of the restricted SNe model only reaches $\sim$ 25\% of the classically expected momentum.  Figure \ref{fig:Normalized} also shows models that assume non-rotating stars. Rotation increases the stellar wind power, but it also extends stellar evolution and therefore effectively shifts the maximum SN progenitor to slightly lower masses.  The overall effect is a slight reduction in total energy injection.

\pagebreak
\section{Additional Mechanisms}

\subsection{Accretion Driven Feedback}\label{sec:accretion}
\citet{Justham2012} also noted the expected delay in SN feedback due to the most massive stars collapsing directly to black holes. They therefore suggest that mechanical feedback from high-mass X-ray binaries (HMXBs) generated by some of these black holes could potentially be important at times before SN feedback dominates.  However, the binary population synthesis models of \citet{Rappaport2005}, predict that the total expected mechanical luminosity from HMXBs at ages $< 10^7$ yr is only on the order of $\log L_{\rm mech}/\ergs\sim 38 - 39$ for a $10^6\ \msun$ stellar populations. 
The left panel of Figure~\ref{fig:Feedback} shows mechanical feedback from the Rappaport et al. models overplotted, normalized to the $10^4$ core-collapse events in our Starburst99 models.  
We assume that the X-ray luminosity
$L_X \sim L_{\rm mech}$ is on the order of the Eddington luminosity $L_{\rm Edd}$ \citep[e.g.,][]{Pinto2022, Justham2012}.  Predictions \citep[e.g.,][]{King2016} and observations \citep[e.g.,][]{Kosec2018, Tao2019} suggest that $L_{\rm mech}$ could also be $10 - 100\times L_{\rm Edd}$, while on the other hand, $L_{\rm Edd}$ may be a substantial upper limit to $L_X$ \citep[e.g.,][]{Kosec2018}.
The figure shows models where $L_{\rm mech} = L_{\rm Edd}$ (S) and $10 L_{\rm Edd}$ (B).
 
We see that HMXB feedback could increase $L_{\rm mech}$ in the minima of the troughs at age 5 -- 10 Myr seen in Figure~\ref{fig:Feedback}, and the effect also would be slightly enhanced at low metallicity \citep[e.g.,][]{Renzo2019}. 
It is important to note that both HMXB feedback and SNe are stochastic.
But in general, the order-of-magnitude reduction in $L_{\rm mech}$ preceding the onset of SNe remains.  

Individual ultraluminous X-ray sources (ULXs) have larger $\log L_X/\ergs \sim 39 - 40$, which is $L_{\rm Edd}$ of black holes with masses up to  $\sim 100\ \msun$.  Thus, individual objects could be important, but due to their stochasticity and unknown $L_{\rm mech}$, it is unclear whether they are an important systematic effect \citep{Justham2012}.  

\subsection{Cluster Evaporation} \label{sec:Evaporation} 

Another mechanism that affects the output of mechanical feedback from a given cluster is the removal of stars by dynamical processes. Dynamical evolution can remove massive stars before they explode as SNe, thereby further reducing mechanical feedback. Massive stars are more likely to be ejected, and if ejected early and fast enough, their stellar winds and possible SNe will be located outside the cluster's sphere of influence, nominally the superbubble or outflow 
radius.
Dynamical
processes which remove stars are dependent on several key initial conditions \citep[see e.g., review by][]{PortegiesZwart2010}, including the initial parameters for cluster density, star formation efficiency, mass segregation, cluster mass, and binary population \citep[e.g.,][]{Oh2016, Brinkmann2017}. 

\citet{Oh2016} find that the initial cluster density is the most influential parameter in their study, with the densest
clusters ejecting 50\% of their O stars by age 3 Myr while the least dense ones eject only 4.5\% at 3 Myr. Primordial mass segregation enhances this effect, since such clusters already have a pre-existing dense core of massive stars where most dynamical ejections occur \citep[e.g.,][]{Oh2016}. However, the amount and timescale of gas dispersal significantly affects the stellar density, which in turn strongly affects the stellar ejection rate \citep{Pfalzner2013}. Similarly, \citet{Brinkmann2017} find that clusters with the same initial density can have widely differing bound fractions depending on gas dispersal parameters. {\it Thus, weak early feedback promotes stellar dynamical ejections,} further removing sources of mechanical feedback.

The stellar ejection fraction peaks at moderately massive clusters ($10^{3.5}\ M_\odot$), including for models with different initial radii \citep{Oh2015}. \citet{Brinkmann2017} find that massive clusters ($> 10^5\ \msun$) have an 80\% bound fraction while moderately massive clusters ($5 \times 10^3\  \msun$) have only a 20\% bound fraction for currently expected gas expulsion parameters. As noted above, the most massive stars tend to sink to the cluster center, promoting their ejection.

Numerical simulations by, e.g., \citet{Oh2015,Oh2016,Brinkmann2017} suggest that most massive stars are ejected within the first 5 Myr. Furthermore, the efficiency of dynamical ejections peaks after 1 Myr for all moderately massive, mass segregated models simulated by \citet{Oh2016}. These massive ejected stars must be outside the superbubble radius to not contribute to the mechanical feedback of the cluster. Velocity distributions of \citet{Oh2016} show that $< ~25$\% of ejected O-stars and B-stars will remain bound at 3 Myr. If SNe do not start until an age of 10 Myr, the remaining ejected yet local SN progenitors have another 7 Myr to escape the cluster. Ejected stars could easily travel 50 -- 100 pc in only a few Myr.

Therefore, the overall effect of dynamical ejections is to enhance the delay in the onset of massive-star mechanical feedback at low metallicity. Accounting for dynamical ejections would further decrease, and potentially extend, the weak mechanical feedback at ages $\lesssim 10$ Myr.

\section{Discussion} \label{sec:Discussion} 

\subsection{Evidence and Implications} \label{sec:Implications} 
Thus, at early times $< 10$ Myr in low-$Z$ systems, we expect only very weak mechanical feedback from metal-poor stellar winds in gas-rich environments.  These conditions promote catastrophic cooling \citep[e.g.,][]{Silich2004, Wuensch2008, Krumholz2017}, pressure-confinement \citep[e.g.,][]{Silich2007, Oey2004, Smith2006}, or delayed launching \citep{Danehkar2021} of any adiabatic mechanical feedback or superwinds.  Observational evidence for missing or suppressed superwinds is building; nearby starbursts such as NGC 5253 \citep{Turner2017}, M82 \citep{Smith2006, Westmoquette2013}, and Mrk~71 \citep{Oey2017, Komarova2021} show no evidence of classical, energy-driven superwinds, and instead show giant molecular clouds within a few pc of the super star clusters, with evidence of metal retention in one instance \citep{Westmoquette2013}.  High nebular ionization seen in species like He\thinspace{\sc ii}, {\sc C iv}, and {\sc O vi} observed in such systems may be evidence that superwinds are being suppressed by weak mechanical input and/or catastrophic cooling \citep[e.g.,][]{Danehkar2021, Danehkar2022, Oey2023}. Moreover, some of the most extreme, metal-poor, Green Pea galaxies, a population including many Lyman-continuum emitters \citep{Flury2022}, show the lowest superwind velocities \citep{Jaskot2017}.

If mechanical feedback effectively starts later at low metallicity than at solar metallicity, it should have profound effects on the structure and evolution of starbursts and their host galaxies. In particular, it implies that radiation feedback dominates over mechanical feedback for the first $\sim10$ Myr \citep[e.g.,][]{Freyer2003, Krumholz2009}. Weaker stellar winds and fewer SNe imply increased gas retention and less negative feedback to disrupt ongoing star formation. This effect may be amplified by the negative cycle of poor gas expulsion enhancing stellar densities and thus, stellar ejections, further weakening both radiative and mechanical feedback (Section~\ref{sec:Evaporation}), which thus may be insufficient to suppress ongoing star formation. This leads to an increase in the star formation efficiency, star formation rate, and timescale for star formation in a given region \citep[e.g.,][]{Shima2017}.
It is consistent with the higher star formation efficiencies seen in super star clusters \citep[e.g.,][]{Turner2015,Herrera2017,Oey2017}, and it has been linked to the formation of multiple stellar populations found in globular clusters \citep{Krause2012, Lochhaas2017, Silich2018}. 

Additionally, the standard paradigm for Lyman-continuum (LyC) escape from starburst regions assumes that mechanically driven superwinds clear channels in the ISM allowing LyC to travel without being absorbed \citep[e.g.,][]{Heckman2001}. Our new model for delayed superwinds poses a problem for 
this scenario of
LyC escape, since at age $\sim$ 10 Myr, the emission rate of LyC photons is $> 100\times$ lower than at unevolved times \citep[e.g.,][]{Leitherer2014}. Hence there must be another mechanism to create channels for LyC escape in young, gas-rich conditions. \citet{Jaskot2019} propose that large quantities of retained gas naturally cool and clump in the absence of strong mechanical feedback, and therefore create gaps between the clumps, providing channels for LyC to escape. This scenario is consistent with the ``picket fence'' geometry that is often invoked for LyC emitters \citep{Heckman2001, Heckman2011, RiveraThorsen2017}.
It is also consistent with simulatioms by, e.g., \citet{Rogers2013} and \citet{Dale2013}, who find that on the order of 50\% or more of LyC photons can escape from star-forming giant molecular clouds due to their inhomogeneous, clumpy nature and radiation-dominated feedback.  Moreover, \citet{Kimm2019} find that LyC escape is enhanced in metal-poor clouds, in particular.

Observations of the local starburst Mrk~71-A appear to support this scenario.  This object provides the strongest evidence of catastrophic cooling where adiabatic mechanical feedback is suppressed, as evidenced by the gas kinematics \citep{Komarova2021, Oey2017} and direct observation of strong radiative cooling \citep{Oey2023}.  \citet{Komarova2021} find that LyC radiation drives a fast superwind out to hundreds of pc from the parent SSC in this object.  This strongly suggests that the radiation is able to escape even further beyond this region and is therefore optically thin, despite the suppression of energy-driven mechanical feedback.  Since the SSC is obscured by high-density gas, including molecular gas \citep{Oey2017}, the presence of the LyC-driven wind therefore implies that the radiation is able to escape through gaps in this gas, consistent with the model of \citet{Jaskot2019}.

A more straightforward consequence of delayed mechanical feedback is that the total mechanical energy from a given cluster will be reduced. In Figure \ref{fig:Normalized} we show the total cumulative energy of the restricted SN model relative to the classical solar model and the subsolar, unrestricted SN model. For the first 10 Myr, the energy from the restricted SN model is just 10\% of the classically expected solar model. Around this age, the SNe begin to explode and quickly dominate the total energy. However, while the total mechanical energy in the metal-poor models is lower than at $\zsun$, the reduction is less than a factor of two, and therefore order-of-magnitude estimates for mechanical feedback are still valid.  

However, the fact that it is effectively delayed for 10 Myr may have fundamental consequences. For example, delayed SN feedback may resolve the superbubble growth-rate discrepancy \citep[e.g.,][]{Oey2009}, in which superbubbles around OB associations in the Magellanic Clouds and the Milky Way appear to be too small for their observed stellar populations \citep[e.g.,][]{,Oey1995,Oey1996, Cooper2004,Smith2004}.  This discrepancy could be resolved if the mechanical luminosity was reduced by a factor of ten \citep{Oey1996}. Figure \ref{fig:Feedback} shows that excluding SNe from the first $\sim$ 10 Myr indeed reduces the mechanical power by about the right amount during this early period when stellar winds dominate and O stars are prevalent. The cited studies of objects in the Magellanic Clouds and Milky Way all have ages in this range. Similarly, if a given cluster's mechanical power is overestimated, then superbubble ages are correspondingly underestimated for given observed radii. This is especially true for objects estimated to be $<15$ Myr old.

On a global scale, the porosity of the neutral ISM will be correspondingly reduced if superbubbles are smaller \citep{Oey1997, Clarke2002}.  Simplistically, there will be $\sim$ 40\% less hot gas generated than expected from prior models, modestly affecting the phase balance of the ISM. Interestingly, the observed ISM porosities of Local Group galaxies estimated by \citet{Oey1997} tend to be lower than the predicted values based on the observed {\sc H ii} region luminosity function.
The reduction in hot gas and superwinds may be exacerbated by radiative cooling enhanced by weak feedback \citep[e.g.,][]{Danehkar2022}. 

Our study complements the large body of work simulating the effects of SN feedback on their host galaxies \citep[see, e.g.,][and references therein]{Keller2022b}.  
\citet{Keller2022} specifically explore how SN parameters, including the time of SN onset and the duration of SNe, affect the regulation of star formation and galactic outflows.  They find that longer delay times indeed enhance the star formation efficiency of individual clouds as noted above.  
Based on individualized timing and location of momentum, energy, and chemical enrichment from stars, \citet{Andersson2023} agree that reduced mechanical feedback leads to colder and more fragmented disks and thus much higher star formation rates. 
\citet{Semenov2021} and \citet{Keller2022b} find similar results for the presence or absence of early, pre-SN feedback, which have significant effects on the ISM structure, subsequent star formation, and nature of superwinds.  
\citet{Gutcke2021} find that allowing for variable SN feedback reduces the total energy, and thus dwarf galaxies will expel slightly less gas and metals. They confirm that more spatially distributed SNe inhibit the development of superwind outflows \citep{Clarke2002} and mass loading; this effect is enhanced when progenitor stars are ejected from clusters.

We have shown that all of these effects are especially prevalent in metal-poor, dwarf galaxies, which therefore {\it may explain why dwarf irregular galaxies have such large star-forming complexes and high specific star-formation rates.}
These processes may also be linked to the compactness of blue compact dwarf galaxies and Green Peas, many of which appear to be responsible for Lyman continuum emission \citep[e.g.,][]{Flury2022, Jaskot2019}.
Weak mechanical feedback is also suggested to be a driving factor in star formation observed at cosmic dawn in JWST observations \citep{Dekel2023}.

\subsection{Element Yields and Abundances}

\begin{figure*}
    \includegraphics[width=2.3in]{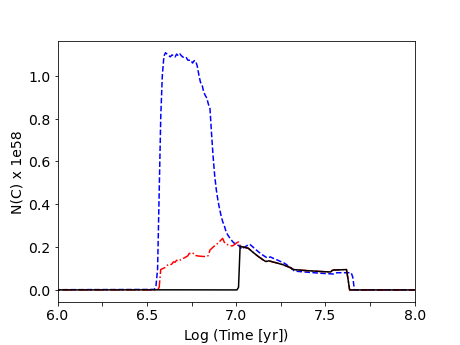}
    \includegraphics[width=2.3in]{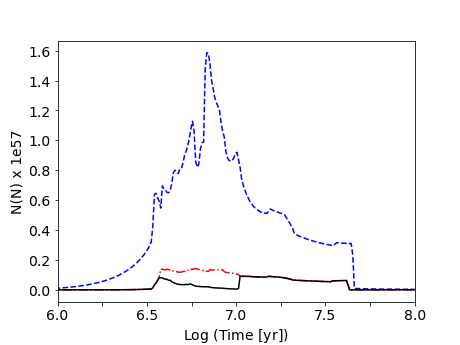}
    \includegraphics[width=2.3in]{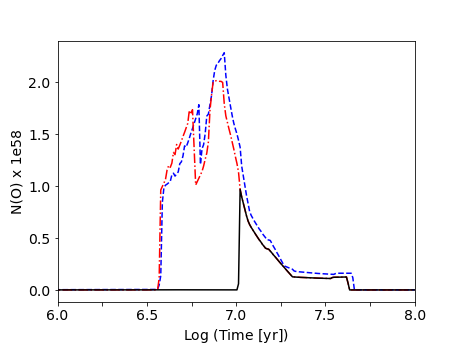}
    \includegraphics[width=2.3in]{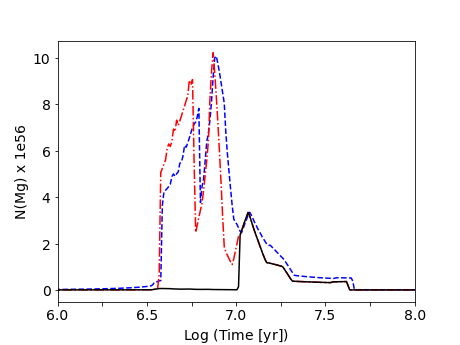}
   \includegraphics[width=2.3in]{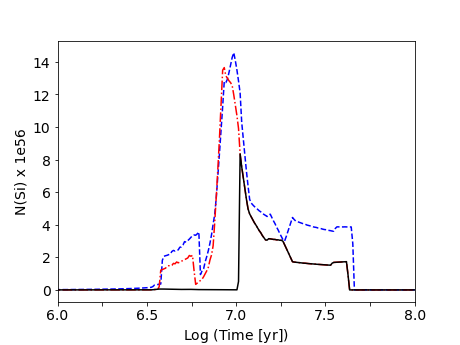}
    \includegraphics[width=2.3in]{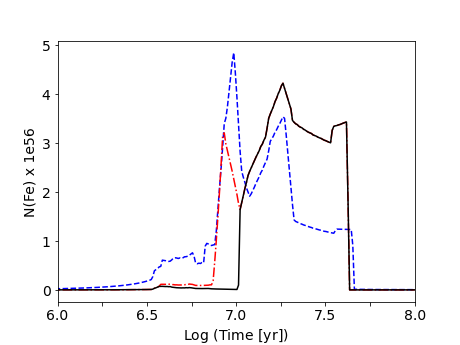}
  \caption{ Production of C, N, O, Mg, Si and Fe by number over time, for the three model populations of $10^6\ \msun$.  
  Line types are as in Figure~\ref{fig:Cumulative}.
  }
  \label{fig:abdsN}
\end{figure*}

\begin{figure}
    \includegraphics[width=3.4in]{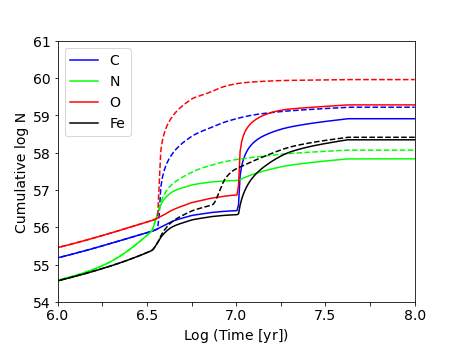}
  \caption{Cumulative production of C, N, O, and Fe for the models at subsolar metallicity. Dashed and solid lines show unrestricted and restricted SN models, respectively.}
  \label{fig:abdsCum}
\end{figure}

\floattable
\begin{deluxetable}{c c c c c c c c c}
\tabletypesize{\scriptsize}
\tablecaption{Element Yields\tablenotemark{a}}
\tablehead{& \colhead{Model} & \colhead{C} & \colhead{N} & \colhead{O\tablenotemark{b}} & \colhead{Mg} & \colhead{Si} & \colhead{S} & \colhead{Fe}}
\startdata
$N$(X, res) / $N$(X, unr)\tablenotemark{c} & Restricted/Unrestricted & 0.492 & 0.582 & 0.209 & 0.207 & 0.459 & 0.498 & 0.856 \\
\tableline
& Solar, Classical & 0.479 & 0.087 & 1.00 & 0.039 & 0.059 & 0.027 & 0.022 \\
X/O \tablenotemark{d} & Subsolar, Unrestricted & 0.181 & 0.013 & 1.00 & 0.038 & 0.045 & 0.018 & 0.028 \\
& Subsolar, Restricted SN & 0.426 & 0.036 & 1.00 & 0.037 & 0.098 & 0.043 & 0.116 \\
\tableline
& Solar Classical & --0.059 & --0.200 & 0.000 & --0.321 & --0.051 & 0.007 & --0.475 \\
$\mathrm{[X/O]}$\tablenotemark{e} & Subsolar, Unrestricted & --0.481 & --1.032 & 0.000 & --0.335 & --0.169 & --0.171 & --0.358 \\
& Subsolar Restricted SN & --0.111 & --0.589 & 0.000 & --0.341 & 0.171 & 0.204 & 0.254
\enddata
\tablenotetext{a}{Values are for a $10^6\ \msun$ Starburst99 population at the time when SN feedback ends at 46.9 Myr.}
\tablenotetext{b}{The total yield of oxygen by number from the subsolar, unrestricted SNe model is 9.14e59; and for the solar, classical model it is 4.84e59.}
\tablenotetext{c}{Ratio of yield produced by subsolar, restricted SNe model relative to subsolar, unrestricted SNe model.}
\tablenotetext{d}{Total yield by number of element X relative to oxygen for each model. }
\tablenotetext{e}{Logarithmic ratio from above, relative to solar photospheric elemental abundances \citep{Asplund2009}.}
\label{tab:abds}
\end{deluxetable}

Under the restricted SN scenario, the SN nucleosynthesis rate will be somewhat lower than generally expected, causing galactic chemical evolution to take place a bit more slowly and with modified element enrichment patterns.  In particular, the $\alpha$/Fe enrichment rate will be slower, since the more massive stars are those that dominate $\alpha$-element yields.  Eliminating SNe from the upper IMF also affects the relative element yields.  
Figure~\ref{fig:abdsN} shows the production of C, N, O, Mg, Si, and Fe during the modeled timeframe for our three models.  
The Starburst99 SN yields are from \citet{Woosley1995}, and stellar wind yields are from the Geneva evolution models.
Table~\ref{tab:abds} gives the total yields for all elements calculated by Starburst99, relative to O, for our model populations.
The first row shows the yield reduction for the subsolar-metallicity, restricted SNe model relative to the unrestricted model where all massive stars explode.  
Additionally, Figure~\ref{fig:abdsCum} shows the cumulative production of C, N, O, and Fe at subsolar metallicity, comparing the restricted and unrestricted SN models.

The data show that restricting the range of SN progenitors causes significant changes in production among the shown elements, and these may have noticeable effects on abundance patterns of $\alpha$-elements and other species due to massive stars.  In particular, O and Mg are produced at only 20\% the rate for full-IMF SNe (Table~\ref{tab:abds}), since they are disproportionately generated by the most massive stars. This is more than a factor of 2 below the typical reduction factors for other elements.  Interestingly, standard galactic chemical evolution models for the Milky Way slightly overpredict the solar O abundance relative to other species \citep[see, e.g., review by ][]{Prantzos2008}. An underproduction of O might also be relevant to the enhanced N/O ratios seen for young systems dominated by primary N evolution \citep[see, e.g., review by][]{Maiolino2019}, and similar enhancements in C/O at low metallicity are suggested to be linked to the same processes dominating early N evolution \citep[e.g.,][]{Maiolino2019, Berg2016}.

On the other hand, the Fe yield changes the least, with a production rate of 86\% that of unrestricted SNe.  Since Fe production is dominated by Type Ia SNe, massive-star production effects on its long-term chemical abundance patterns is further minimized.  We also note that the effects of suppressed feedback are likely unimportant for the interpretation of abundances in extremely metal-poor (EMP) stars.  Although these are dominated by core-collapse SN production, the interpretation is that they reflect the products of individual SN events \citep[e.g.,][]{Frebel2015}, and therefore the collective production of a young population, as modeled here, is not relevant.
Since the number of SNe is dominated by the lower-mass range of progenitors according to the IMF, the effect of changing the maximum SN progenitor mass does not have a strong effect on galactic chemical evolution.  As noted above, it results in changes on the order of 2 -- 3 to some element yields, which is modest relative to evolutionary abundance patterns.  However, as shown in Table~\ref{tab:abds} and Figure~\ref{fig:abdsCum}, these effects may still be significant.  More comprehensive investigation, including the production of other elements, is needed to fully understand the effect of restricting the range of SN progenitors.

\subsection{Uncertainties} \label{sec:Uncertainties} 

The scenario we have presented is subject to a number of uncertainties that affect the timescale for the onset of SN feedback and magnitude of preceding stellar wind feedback. The main source of uncertainty centers on which stars successfully explode as core-collapse SNe. While there is a consensus that most metal-poor high-mass stars collapse directly to BHs without SN explosions \citep[e.g.,][]{Zhang2008, O'Connor2011,Sukhbold2016,Muller2016,Patton2020}, exactly which progenitor masses explode and why has yet to be definitively determined. \citet{O'Connor2011} find that a progenitor's core compactness at bounce determines its final fate. At low metallicities, this parameter is what causes the failure of stars more massive than $\sim$20 $\msun$ to explode. \citet{Ertl2016} find a related explosion criterion that produces similar results, and finds that only $\sim$10\% of stars more massive than 20 $\msun$ generate SNe \citep{Sukhbold2016}. 
\citet{Keller2022} also stress the importance of the {\it minimum}-mass SN progenitor, which strongly influences the total energy and momentum injected by a given IMF.
The stellar structure, explosion mechanics, and progenitor mass range are also important for nucleosynthetic yields \citep[e.g.,][]{Prantzos2008}.

There are also certain mass ranges that tend to explode. \citet{Sukhbold2016} find that SN progenitors less massive than 15 $\msun$ explode easily, while those in the ranges 15 -- 23 $\msun$, 27 -- 30 $\msun$, or  $>35\ \msun$ explode rarely. For their one-dimensional core-collapse models, the vast majority of SN progenitors having masses $> 20\ \msun$ do not explode. We adopted this value as a threshold for our simple parameterization above that assumes all stars with higher masses collapse directly into black holes. On the other hand, 3-D models show higher turbulence and convection, which may enhance explodability \citep{Fields2021}.

Moreover, most models of explodability agree that the masses of exploding stars are not a continuous distribution.
Although not fully understood yet, the explosion mechanism of core-collapse supernovae is thought to be inherently stochastic \citep[e.g.,][]{Cardall2016}. This introduces variance into the explosion timescale, yields, and SN mass threshold, so the start of CCSN feedback will be more gradual than modeled in Figure \ref{fig:Feedback}, and some SNe are expected to occur before 10 Myr. 

In addition, very massive stars are predicted to end their lives in rare, energetic ($< 10^{53}$ erg) pair-instability supernovae (PISNe). \citet{Heger2002} find that stars with helium core masses 64 -- 133 $\msun$ produce PISNe. This corresponds to initial masses of ~140 -- 260 $\msun$, of which we have $\sim100$ in our 10$^{6}$ $\msun$ cluster \citep{Kasen2011, Heger2003, Hirschi2017}. If these predictions are correct, this would produce significant mechanical luminosity from the cluster at early ages. However, a single PISN event has yet to be reliably identified, even with $>1,000$ SNe detected per year and a relative event rate of PISNe to CCSNe of 1\%.  \citet{Takahashi2018} therefore suggests that there may be fewer PISNe than expected due to shell convection occurring earlier in the core carbon-burning phase, whereby PISN progenitors can avoid pair-creation instability. 

Moreover, there are still significant uncertainties in stellar wind parameters, especially at low metallicity. Uncertain mass-loss rates and wind velocities are a major problem in creating accurate stellar evolution models and population synthesis models. The most commonly used formula for mass loss is that of \citet{Vink2001}, which is now believed to be an overestimate. Accounting for clumping and wind hydrodynamics has been shown to lower theoretical mass-loss rates by at least factor of 2 -- 3 \citep{Bouret2012,Surlan2013,Puls2008}. However, UV spectral observations of O-stars in the massive young SMC regions NGC 346 and SMC-SGS 1 show even more dramatic mass-loss rate reductions by factors closer to an order of magnitude or more \citep{Rickard2022, Ramachandran2019} from the theoretical values of \citet{Vink2001}. 

Another important uncertainty originates from the omission of binary stars from our analysis. Accounting for massive binary evolution will extend the lives of binary mass gainers and therefore add SNe at later ages.  It will also increase the numbers
of HMXBs and ULXs, and their feedback. 
We suggested above (Section~\ref{sec:accretion}) that
neither of these effects will significantly change the weak mechanical feedback at early times, but further study is needed to clarify their contribution.
Massive, interacting binaries lead to increased LyC emission and stellar winds from stripped stars, possibly increasing both radiative and mechanical feedback. However binaries also drive ejection of massive stars from their clusters, which is likely to offset this effect.

Furthermore, binarity is now believed to dominate massive star evolution.  
Binaries dominate dynamical processes leading to stellar ejections, and further work is needed to quantify the loss of mechanical feedback due to this effect, which, as we argued above, is likely substantial.  Binary
mass transfer will greatly increase rotation velocities of mass gainers, while also greatly reducing the masses of donors.  The SB99 models have only two stellar rotational velocities, 0 and 40\% of the ZAMS break-up velocity. As is evident in Figure \ref{fig:Normalized}, faster rotating stars further delay the onset of mechanical feedback. Binary mass transfer could cause much higher rotation velocities, which would thus further shift the start of strong mechanical feedback to somewhat later times.  Binary mergers could also affect the mechanics and likelihood of SN explosions and their nucleosynthetic yields.

\section{Conclusion} \label{sec:Conclusion} 

In summary, massive-star feedback at low metallicity differs greatly from the classical paradigm at solar metallicity. At low metallicity, stellar winds are weak, and the more massive stars fail to explode as core-collapse SNe. We have compared the evolution of mechanical luminosity and momentum injection for low metallicity models to those predicted by the classical, $\zsun$ feedback paradigm. Our restricted SNe model limits SNe to progenitor masses $< 20\ \msun$, as compared to the classical, unrestricted limit. We find that low metallicity Starburst99 models predict that mechanical feedback is effectively delayed by roughly 10 Myr.

We additionally discussed the contribution of HMXBs to our low metallicity, delayed feedback model. We found that accretion-driven feedback could slightly increase the mechanical luminosity of the cluster only between the ages 5-10 Myr, but the order-of-magnitude reduction during this time remains. 

Furthermore, dynamical mechanisms remove stars from clusters, reducing their mechanical feedback contribution. This is often an overlooked, but likely significant process. The effect is enhanced for clusters that are mass-segregated, dense, have high binary fractions, and have moderate masses \citep{Oh2015}. Such clusters at low metallicity will deviate the most from the classical mechanical feedback paradigm. For example, the moderately massive and densest cluster modeled by \citet{Oh2016} ejected 50\% of its O-stars by 3 Myr. A potentially important process that remains to be explored is that {\it gas retention due to weak feedback likely promotes ejection of OB stars, exacerbating the effect.}  This occurs because gas retention
further promotes mass segregation and high cluster densities. Accounting for the dynamical mechanisms that remove stars will thus further decrease the expected total mechanical feedback of a given cluster. 

A delay in the onset of mechanical feedback implies that radiation dominates over mechanical feedback at early ages. This corresponds to the scenario where mechanically driven superwinds are suppressed by catastrophic cooling or pressure confinement, as has been observed in several local starburst systems. This has a variety of important implications. There is increased gas retention, which increases rate and timescale of star formation. This is consistent with the higher star formation efficiencies in super star clusters and may be linked to the multiple stellar populations in globular clusters. 
{\it Delayed feedback may also offer a simple explanation for why metal-poor, dwarf irregular galaxies have such large star-forming complexes.} 
Retention of dense gas near super star clusters likely leads to clumping, thereby generating the picket-fence geometry that may be conducive to LyC escape through inter-clump regions. 

Overestimated early mechanical luminosity can explain why superbubbles are often found to be too small for their observed parent stellar populations. Similarly, superbubble ages that are inferred from the observed radii and stellar populations are therefore too young. 
The cumulative, large-scale effect of reduced initial mechanical luminosity is to reduce the total integrated energy and production of hot gas from clusters by $\sim$40\% and to reduce the total momentum injected by 75\% relative to the classical model. This correspondingly, modestly modifies the phase balance of the ISM and CGM. It also has a modest effect on galactic chemical evolution by slowing the $\alpha$/Fe evolution rate, and slightly modifying abundance patterns. In particular, the production of O is reduced by a factor $\gtrsim 2$ relative to other elements.

These effects are subject to a number of uncertainties, in particular, involving the stellar progenitors that successfully explode as core-collapse SNe; and the myriad effects of binary evolution, which is unaccounted for in our models. However, this model of delayed mechanical feedback cohesively explains many observations for metal-poor star-forming regions and starburst galaxies, including the character of star formation in metal-poor, dwarf galaxies and starbursts. Since the associated feedback processes play a key role in our universe, both at early epochs and the present, this effect is broadly relevant and worth deeper investigation.

\begin{acknowledgements}
We wish to thank Carl Fields, Edmund Hodges-Kluck, Anne Jaskot, Evan Kirby, Lena Komarova, Claus Leitherer, and Vadim Semenov for helpful comments and discussions. Additionally, we thank the anonymous referee for valuable suggestions which improved the depth of the paper. This work was supported by NASA HST-GO-16261 and the University of Michigan.
\end{acknowledgements}


\bibliography{MechFeedback.bib}{}

\begin{thebibliography}{}
\expandafter\ifx\csname natexlab\endcsname\relax\def\natexlab#1{#1}\fi
\providecommand{\url}[1]{\href{#1}{#1}}
\providecommand{\dodoi}[1]{doi:~\href{http://doi.org/#1}{\nolinkurl{#1}}}
\providecommand{\doeprint}[1]{\href{http://ascl.net/#1}{\nolinkurl{http://ascl.net/#1}}}
\providecommand{\doarXiv}[1]{\href{https://arxiv.org/abs/#1}{\nolinkurl{https://arxiv.org/abs/#1}}}

\bibitem[{{Andersson} {et~al.}(2023){Andersson}, {Agertz}, {Renaud}, \& {Teyssier}}]{Andersson2023}
{Andersson}, E.~P., {Agertz}, O., {Renaud}, F., \& {Teyssier}, R. 2023, \mnras, 521, 2196, \dodoi{10.1093/mnras/stad692}

\bibitem[{{Asplund} {et~al.}(2009){Asplund}, {Grevesse}, {Sauval}, \& {Scott}}]{Asplund2009}
{Asplund}, M., {Grevesse}, N., {Sauval}, A.~J., \& {Scott}, P. 2009, \araa, 47, 481, \dodoi{10.1146/annurev.astro.46.060407.145222}

\bibitem[{{Berg} {et~al.}(2016){Berg}, {Skillman}, {Henry}, {Erb}, \& {Carigi}}]{Berg2016}
{Berg}, D.~A., {Skillman}, E.~D., {Henry}, R. B.~C., {Erb}, D.~K., \& {Carigi}, L. 2016, \apj, 827, 126, \dodoi{10.3847/0004-637X/827/2/126}

\bibitem[{{Bj{\"o}rklund} {et~al.}(2022){Bj{\"o}rklund}, {Sundqvist}, {Singh}, {Puls}, \& {Najarro}}]{Bjorklund2022}
{Bj{\"o}rklund}, R., {Sundqvist}, J.~O., {Singh}, S.~M., {Puls}, J., \& {Najarro}, F. 2022, arXiv e-prints, arXiv:2203.08218.
\newblock \doarXiv{2203.08218}

\bibitem[{{Bouret} {et~al.}(2012){Bouret}, {Hillier}, {Lanz}, \& {Fullerton}}]{Bouret2012}
{Bouret}, J.~C., {Hillier}, D.~J., {Lanz}, T., \& {Fullerton}, A.~W. 2012, \aap, 544, A67, \dodoi{10.1051/0004-6361/201118594}

\bibitem[{{Brinkmann} {et~al.}(2017){Brinkmann}, {Banerjee}, {Motwani}, \& {Kroupa}}]{Brinkmann2017}
{Brinkmann}, N., {Banerjee}, S., {Motwani}, B., \& {Kroupa}, P. 2017, \aap, 600, A49, \dodoi{10.1051/0004-6361/201629312}

\bibitem[{{Cardall} \& {Budiardja}(2016)}]{Cardall2016}
{Cardall}, C., \& {Budiardja}, R. 2016, in APS Meeting Abstracts, Vol. 2016, APS April Meeting Abstracts, M13.002

\bibitem[{{Clarke} \& {Oey}(2002)}]{Clarke2002}
{Clarke}, C., \& {Oey}, M.~S. 2002, \mnras, 337, 1299, \dodoi{10.1046/j.1365-8711.2002.05976.x}

\bibitem[{{Cooper} {et~al.}(2004){Cooper}, {Guerrero}, {Chu}, {Chen}, \& {Dunne}}]{Cooper2004}
{Cooper}, R.~L., {Guerrero}, M.~A., {Chu}, Y.-H., {Chen}, C. H.~R., \& {Dunne}, B.~C. 2004, \apj, 605, 751, \dodoi{10.1086/382501}

\bibitem[{{Dale} {et~al.}(2013){Dale}, {Ercolano}, \& {Bonnell}}]{Dale2013}
{Dale}, J.~E., {Ercolano}, B., \& {Bonnell}, I.~A. 2013, \mnras, 430, 234, \dodoi{10.1093/mnras/sts592}

\bibitem[{{Danehkar} {et~al.}(2021){Danehkar}, {Oey}, \& {Gray}}]{Danehkar2021}
{Danehkar}, A., {Oey}, M.~S., \& {Gray}, W.~J. 2021, \apj, 921, 91, \dodoi{10.3847/1538-4357/ac1a76}

\bibitem[{{Danehkar} {et~al.}(2022){Danehkar}, {Oey}, \& {Gray}}]{Danehkar2022}
---. 2022, \apj, 937, 68, \dodoi{10.3847/1538-4357/ac8cec}

\bibitem[{{Dekel} {et~al.}(2023){Dekel}, {Sarkar}, {Birnboim}, {Mandelker}, \& {Li}}]{Dekel2023}
{Dekel}, A., {Sarkar}, K.~C., {Birnboim}, Y., {Mandelker}, N., \& {Li}, Z. 2023, \mnras, 523, 3201, \dodoi{10.1093/mnras/stad1557}

\bibitem[{{Ekstr{\"o}m} {et~al.}(2012){Ekstr{\"o}m}, {Georgy}, {Eggenberger}, {Meynet}, {Mowlavi}, {Wyttenbach}, {Granada}, {Decressin}, {Hirschi}, {Frischknecht}, {Charbonnel}, \& {Maeder}}]{Ekstr2012}
{Ekstr{\"o}m}, S., {Georgy}, C., {Eggenberger}, P., {et~al.} 2012, \aap, 537, A146, \dodoi{10.1051/0004-6361/201117751}

\bibitem[{{Ertl} {et~al.}(2016){Ertl}, {Janka}, {Woosley}, {Sukhbold}, \& {Ugliano}}]{Ertl2016}
{Ertl}, T., {Janka}, H.~T., {Woosley}, S.~E., {Sukhbold}, T., \& {Ugliano}, M. 2016, \apj, 818, 124, \dodoi{10.3847/0004-637X/818/2/124}

\bibitem[{{Fields} \& {Couch}(2021)}]{Fields2021}
{Fields}, C.~E., \& {Couch}, S.~M. 2021, \apj, 921, 28, \dodoi{10.3847/1538-4357/ac24fb}

\bibitem[{Flury {et~al.}(2022)Flury, Jaskot, Ferguson, Worseck, Makan, Chisholm, Saldana-Lopez, Schaerer, McCandliss, Wang, Ford, Heckman, Ji, Giavalisco, Amorin, Atek, Blaizot, Borthakur, Carr, Castellano, Cristiani, De~Barros, Dickinson, Finkelstein, Fleming, Fontanot, Garel, Grazian, Hayes, Henry, Mauerhofer, Micheva, Oey, Ostlin, Papovich, Pentericci, Ravindranath, Rosdahl, Rutkowski, Santini, Scarlata, Teplitz, Thuan, Trebitsch, Vanzella, Verhamme, \& Xu}]{Flury2022}
Flury, S.~R., Jaskot, A.~E., Ferguson, H.~C., {et~al.} 2022, \apjs, 260, 1, \dodoi{10.3847/1538-4365/ac5331}

\bibitem[{{Frebel} \& {Norris}(2015)}]{Frebel2015}
{Frebel}, A., \& {Norris}, J.~E. 2015, \araa, 53, 631, \dodoi{10.1146/annurev-astro-082214-122423}

\bibitem[{{Freyer} {et~al.}(2003){Freyer}, {Hensler}, \& {Yorke}}]{Freyer2003}
{Freyer}, T., {Hensler}, G., \& {Yorke}, H.~W. 2003, \apj, 594, 888, \dodoi{10.1086/376937}

\bibitem[{{Georgy} {et~al.}(2013){Georgy}, {Ekstr{\"o}m}, {Eggenberger}, {Meynet}, {Haemmerl{\'e}}, {Maeder}, {Granada}, {Groh}, {Hirschi}, {Mowlavi}, {Yusof}, {Charbonnel}, {Decressin}, \& {Barblan}}]{Georgy2013}
{Georgy}, C., {Ekstr{\"o}m}, S., {Eggenberger}, P., {et~al.} 2013, \aap, 558, A103, \dodoi{10.1051/0004-6361/201322178}

\bibitem[{{Gormaz-Matamala} {et~al.}(2022){Gormaz-Matamala}, {Cur{\'e}}, {Meynet}, {Cuadra}, {Groh}, \& {Murphy}}]{Gormaz2022}
{Gormaz-Matamala}, A.~C., {Cur{\'e}}, M., {Meynet}, G., {et~al.} 2022, arXiv e-prints, arXiv:2207.04786.
\newblock \doarXiv{2207.04786}

\bibitem[{{Gutcke} {et~al.}(2021){Gutcke}, {Pakmor}, {Naab}, \& {Springel}}]{Gutcke2021}
{Gutcke}, T.~A., {Pakmor}, R., {Naab}, T., \& {Springel}, V. 2021, \mnras, 501, 5597, \dodoi{10.1093/mnras/staa3875}

\bibitem[{{Heckman} {et~al.}(2001){Heckman}, {Sembach}, {Meurer}, {Leitherer}, {Calzetti}, \& {Martin}}]{Heckman2001}
{Heckman}, T.~M., {Sembach}, K.~R., {Meurer}, G.~R., {et~al.} 2001, \apj, 558, 56, \dodoi{10.1086/322475}

\bibitem[{{Heckman} {et~al.}(2011){Heckman}, {Borthakur}, {Overzier}, {Kauffmann}, {Basu-Zych}, {Leitherer}, {Sembach}, {Martin}, {Rich}, {Schiminovich}, \& {Seibert}}]{Heckman2011}
{Heckman}, T.~M., {Borthakur}, S., {Overzier}, R., {et~al.} 2011, \apj, 730, 5, \dodoi{10.1088/0004-637X/730/1/5}

\bibitem[{{Heger} {et~al.}(2003){Heger}, {Fryer}, {Woosley}, {Langer}, \& {Hartmann}}]{Heger2003}
{Heger}, A., {Fryer}, C.~L., {Woosley}, S.~E., {Langer}, N., \& {Hartmann}, D.~H. 2003, \apj, 591, 288, \dodoi{10.1086/375341}

\bibitem[{{Heger} \& {Woosley}(2002)}]{Heger2002}
{Heger}, A., \& {Woosley}, S.~E. 2002, \apj, 567, 532, \dodoi{10.1086/338487}

\bibitem[{{Herrera} \& {Boulanger}(2017)}]{Herrera2017}
{Herrera}, C.~N., \& {Boulanger}, F. 2017, \aap, 600, A139, \dodoi{10.1051/0004-6361/201628454}

\bibitem[{{Hirschi}(2017)}]{Hirschi2017}
{Hirschi}, R. 2017, in Handbook of Supernovae, ed. A.~W. {Alsabti} \& P.~{Murdin}, 567, \dodoi{10.1007/978-3-319-21846-5_120}

\bibitem[{{Jaskot} {et~al.}(2019){Jaskot}, {Dowd}, {Oey}, {Scarlata}, \& {McKinney}}]{Jaskot2019}
{Jaskot}, A.~E., {Dowd}, T., {Oey}, M.~S., {Scarlata}, C., \& {McKinney}, J. 2019, \apj, 885, 96, \dodoi{10.3847/1538-4357/ab3d3b}

\bibitem[{{Jaskot} {et~al.}(2017){Jaskot}, {Oey}, {Scarlata}, \& {Dowd}}]{Jaskot2017}
{Jaskot}, A.~E., {Oey}, M.~S., {Scarlata}, C., \& {Dowd}, T. 2017, \apjl, 851, L9, \dodoi{10.3847/2041-8213/aa9d83}

\bibitem[{{Justham} \& {Schawinski}(2012)}]{Justham2012}
{Justham}, S., \& {Schawinski}, K. 2012, \mnras, 423, 1641, \dodoi{10.1111/j.1365-2966.2012.20985.x}

\bibitem[{{Kasen} {et~al.}(2011){Kasen}, {Woosley}, \& {Heger}}]{Kasen2011}
{Kasen}, D., {Woosley}, S.~E., \& {Heger}, A. 2011, \apj, 734, 102, \dodoi{10.1088/0004-637X/734/2/102}

\bibitem[{{Keller} \& {Kruijssen}(2022)}]{Keller2022}
{Keller}, B.~W., \& {Kruijssen}, J.~M.~D. 2022, \mnras, 512, 199, \dodoi{10.1093/mnras/stac511}

\bibitem[{{Keller} {et~al.}(2022){Keller}, {Kruijssen}, \& {Chevance}}]{Keller2022b}
{Keller}, B.~W., {Kruijssen}, J.~M.~D., \& {Chevance}, M. 2022, \mnras, 514, 5355, \dodoi{10.1093/mnras/stac1607}

\bibitem[{{Keller} {et~al.}(2014){Keller}, {Wadsley}, {Benincasa}, \& {Couchman}}]{Keller2014}
{Keller}, B.~W., {Wadsley}, J., {Benincasa}, S.~M., \& {Couchman}, H.~M.~P. 2014, \mnras, 442, 3013, \dodoi{10.1093/mnras/stu1058}

\bibitem[{{Kimm} {et~al.}(2019){Kimm}, {Blaizot}, {Garel}, {Michel-Dansac}, {Katz}, {Rosdahl}, {Verhamme}, \& {Haehnelt}}]{Kimm2019}
{Kimm}, T., {Blaizot}, J., {Garel}, T., {et~al.} 2019, \mnras, 486, 2215, \dodoi{10.1093/mnras/stz989}

\bibitem[{{King} \& {Muldrew}(2016)}]{King2016}
{King}, A., \& {Muldrew}, S.~I. 2016, \mnras, 455, 1211, \dodoi{10.1093/mnras/stv2347}

\bibitem[{Komarova {et~al.}(2021)Komarova, Oey, Krumholz, Silich, Kumari, \& James}]{Komarova2021}
Komarova, L., Oey, M.~S., Krumholz, M.~R., {et~al.} 2021, \apjl, 920, L46, \dodoi{10.3847/2041-8213/ac2c09}

\bibitem[{{Kosec} {et~al.}(2018){Kosec}, {Pinto}, {Walton}, {Fabian}, {Bachetti}, {Brightman}, {F{\"u}rst}, \& {Grefenstette}}]{Kosec2018}
{Kosec}, P., {Pinto}, C., {Walton}, D.~J., {et~al.} 2018, \mnras, 479, 3978, \dodoi{10.1093/mnras/sty1626}

\bibitem[{{Krause} {et~al.}(2012){Krause}, {Charbonnel}, {Decressin}, {Meynet}, {Prantzos}, \& {Diehl}}]{Krause2012}
{Krause}, M., {Charbonnel}, C., {Decressin}, T., {et~al.} 2012, \aap, 546, L5, \dodoi{10.1051/0004-6361/201220244}

\bibitem[{{Krumholz} \& {Matzner}(2009)}]{Krumholz2009}
{Krumholz}, M.~R., \& {Matzner}, C.~D. 2009, \apj, 703, 1352, \dodoi{10.1088/0004-637X/703/2/1352}

\bibitem[{{Krumholz} {et~al.}(2017){Krumholz}, {Thompson}, {Ostriker}, \& {Martin}}]{Krumholz2017}
{Krumholz}, M.~R., {Thompson}, T.~A., {Ostriker}, E.~C., \& {Martin}, C.~L. 2017, \mnras, 471, 4061, \dodoi{10.1093/mnras/stx1882}

\bibitem[{{Leitherer} {et~al.}(2014){Leitherer}, {Ekstr{\"o}m}, {Meynet}, {Schaerer}, {Agienko}, \& {Levesque}}]{Leitherer2014}
{Leitherer}, C., {Ekstr{\"o}m}, S., {Meynet}, G., {et~al.} 2014, \apjs, 212, 14, \dodoi{10.1088/0067-0049/212/1/14}

\bibitem[{{Lochhaas} \& {Thompson}(2017)}]{Lochhaas2017}
{Lochhaas}, C., \& {Thompson}, T.~A. 2017, \mnras, 470, 977, \dodoi{10.1093/mnras/stx1289}

\bibitem[{{Maiolino} \& {Mannucci}(2019)}]{Maiolino2019}
{Maiolino}, R., \& {Mannucci}, F. 2019, \aapr, 27, 3, \dodoi{10.1007/s00159-018-0112-2}

\bibitem[{{Meynet} {et~al.}(1994){Meynet}, {Maeder}, {Schaller}, {Schaerer}, \& {Charbonnel}}]{Meynet1994}
{Meynet}, G., {Maeder}, A., {Schaller}, G., {Schaerer}, D., \& {Charbonnel}, C. 1994, \aaps, 103, 97

\bibitem[{{M{\"u}ller} {et~al.}(2016){M{\"u}ller}, {Heger}, {Liptai}, \& {Cameron}}]{Muller2016}
{M{\"u}ller}, B., {Heger}, A., {Liptai}, D., \& {Cameron}, J.~B. 2016, \mnras, 460, 742, \dodoi{10.1093/mnras/stw1083}

\bibitem[{{O'Connor} \& {Ott}(2011)}]{O'Connor2011}
{O'Connor}, E., \& {Ott}, C.~D. 2011, \apj, 730, 70, \dodoi{10.1088/0004-637X/730/2/70}

\bibitem[{{Oey}(1996)}]{Oey1996}
{Oey}, M.~S. 1996, \apj, 467, 666, \dodoi{10.1086/177642}

\bibitem[{{Oey}(2009)}]{Oey2009}
{Oey}, M.~S. 2009, in American Institute of Physics Conference Series, Vol. 1156, The Local Bubble and Beyond II, ed. R.~K. {Smith}, S.~L. {Snowden}, \& K.~D. {Kuntz}, 295--304, \dodoi{10.1063/1.3211829}

\bibitem[{{Oey} \& {Clarke}(1997)}]{Oey1997}
{Oey}, M.~S., \& {Clarke}, C.~J. 1997, \mnras, 289, 570, \dodoi{10.1093/mnras/289.3.570}

\bibitem[{{Oey} \& {Garc{\'\i}a-Segura}(2004)}]{Oey2004}
{Oey}, M.~S., \& {Garc{\'\i}a-Segura}, G. 2004, \apj, 613, 302, \dodoi{10.1086/421483}

\bibitem[{{Oey} {et~al.}(2017){Oey}, {Herrera}, {Silich}, {Reiter}, {James}, {Jaskot}, \& {Micheva}}]{Oey2017}
{Oey}, M.~S., {Herrera}, C.~N., {Silich}, S., {et~al.} 2017, \apjl, 849, L1, \dodoi{10.3847/2041-8213/aa9215}

\bibitem[{{Oey} {et~al.}(2005){Oey}, {Watson}, {Kern}, \& {Walth}}]{Oey2005}
{Oey}, M.~S., {Watson}, A.~M., {Kern}, K., \& {Walth}, G.~L. 2005, \aj, 129, 393, \dodoi{10.1086/426333}

\bibitem[{{Oey} {et~al.}(2023){Oey}, {Sawant}, {Melinder}, {Smith}, {Leitherer}, {Silich}, {Calzetti}, {Chu}, {Hayes}, {James}, {Jaskot}, {Micheva}, {\"Ostlin}, \& {Reiter}}]{Oey2023}
{Oey}, M.~S., {Sawant}, A.~N., {Melinder}, J., {et~al.} 2023, in Astronomy in Focus, Focus Meeting 4, IAU General Assembly XXXI, Vol.~31, Astronomy in Focus, Focus Meeting 4, ed. J.~{Espinosa}, in press

\bibitem[{{Oey}(1995)}]{Oey1995}
{Oey}, M. S.-L. 1995, PhD thesis, University of Arizona

\bibitem[{{Oh} \& {Kroupa}(2016)}]{Oh2016}
{Oh}, S., \& {Kroupa}, P. 2016, \aap, 590, A107, \dodoi{10.1051/0004-6361/201628233}

\bibitem[{{Oh} {et~al.}(2015){Oh}, {Kroupa}, \& {Pflamm-Altenburg}}]{Oh2015}
{Oh}, S., {Kroupa}, P., \& {Pflamm-Altenburg}, J. 2015, \apj, 805, 92, \dodoi{10.1088/0004-637X/805/2/92}

\bibitem[{{Patton} \& {Sukhbold}(2020)}]{Patton2020}
{Patton}, R.~A., \& {Sukhbold}, T. 2020, \mnras, 499, 2803, \dodoi{10.1093/mnras/staa3029}

\bibitem[{{Pfalzner} \& {Kaczmarek}(2013)}]{Pfalzner2013}
{Pfalzner}, S., \& {Kaczmarek}, T. 2013, \aap, 555, A135, \dodoi{10.1051/0004-6361/201321362}

\bibitem[{{Pinto} \& {Kosec}(2022)}]{Pinto2022}
{Pinto}, C., \& {Kosec}, P. 2022, arXiv e-prints, arXiv:2211.00014.
\newblock \doarXiv{2211.00014}

\bibitem[{Portegies~Zwart {et~al.}(2010)Portegies~Zwart, McMillan, \& Gieles}]{PortegiesZwart2010}
Portegies~Zwart, S.~F., McMillan, S.~L., \& Gieles, M. 2010, Annual Review of Astronomy and Astrophysics, 48, 431, \dodoi{10.1146/annurev-astro-081309-130834}

\bibitem[{{Prantzos}(2008)}]{Prantzos2008}
{Prantzos}, N. 2008, in EAS Publications Series, Vol.~32, EAS Publications Series, ed. C.~{Charbonnel} \& J.~P. {Zahn}, 311--356, \dodoi{10.1051/eas:0832009}

\bibitem[{{Puls} {et~al.}(2008){Puls}, {Vink}, \& {Najarro}}]{Puls2008}
{Puls}, J., {Vink}, J.~S., \& {Najarro}, F. 2008, \aapr, 16, 209, \dodoi{10.1007/s00159-008-0015-8}

\bibitem[{{Ramachandran} {et~al.}(2019){Ramachandran}, {Hamann}, {Oskinova}, {Gallagher}, {Hainich}, {Shenar}, {Sander}, {Todt}, \& {Fulmer}}]{Ramachandran2019}
{Ramachandran}, V., {Hamann}, W.~R., {Oskinova}, L.~M., {et~al.} 2019, \aap, 625, A104, \dodoi{10.1051/0004-6361/201935365}

\bibitem[{{Rappaport} {et~al.}(2005){Rappaport}, {Podsiadlowski}, \& {Pfahl}}]{Rappaport2005}
{Rappaport}, S.~A., {Podsiadlowski}, P., \& {Pfahl}, E. 2005, \mnras, 356, 401, \dodoi{10.1111/j.1365-2966.2004.08489.x}

\bibitem[{{Renzo} {et~al.}(2019){Renzo}, {Zapartas}, {de Mink}, {G{\"o}tberg}, {Justham}, {Farmer}, {Izzard}, {Toonen}, \& {Sana}}]{Renzo2019}
{Renzo}, M., {Zapartas}, E., {de Mink}, S.~E., {et~al.} 2019, \aap, 624, A66, \dodoi{10.1051/0004-6361/201833297}

\bibitem[{{Rickard} {et~al.}(2022){Rickard}, {Hainich}, {Hamann}, {Oskinova}, {Prinja}, {Ramachandran}, {Pauli}, {Todt}, {Sander}, {Shenar}, {Chu}, \& {Gallagher}}]{Rickard2022}
{Rickard}, M.~J., {Hainich}, R., {Hamann}, W.~R., {et~al.} 2022, \aap, 666, A189, \dodoi{10.1051/0004-6361/202243281}

\bibitem[{{Rivera-Thorsen} {et~al.}(2017){Rivera-Thorsen}, {{\"O}stlin}, {Hayes}, \& {Puschnig}}]{RiveraThorsen2017}
{Rivera-Thorsen}, T.~E., {{\"O}stlin}, G., {Hayes}, M., \& {Puschnig}, J. 2017, \apj, 837, 29, \dodoi{10.3847/1538-4357/aa5d0a}

\bibitem[{{Rogers} \& {Pittard}(2013)}]{Rogers2013}
{Rogers}, H., \& {Pittard}, J.~M. 2013, \mnras, 431, 1337, \dodoi{10.1093/mnras/stt255}

\bibitem[{{Salpeter}(1955)}]{Salpeter1955}
{Salpeter}, E.~E. 1955, \apj, 121, 161, \dodoi{10.1086/145971}

\bibitem[{{Semenov} {et~al.}(2021){Semenov}, {Kravtsov}, \& {Gnedin}}]{Semenov2021}
{Semenov}, V.~A., {Kravtsov}, A.~V., \& {Gnedin}, N.~Y. 2021, \apj, 918, 13, \dodoi{10.3847/1538-4357/ac0a77}

\bibitem[{{Shima} {et~al.}(2017){Shima}, {Tasker}, \& {Habe}}]{Shima2017}
{Shima}, K., {Tasker}, E.~J., \& {Habe}, A. 2017, \mnras, 467, 512, \dodoi{10.1093/mnras/stw3279}

\bibitem[{{Silich} \& {Tenorio-Tagle}(2018)}]{Silich2018}
{Silich}, S., \& {Tenorio-Tagle}, G. 2018, \mnras, 478, 5112, \dodoi{10.1093/mnras/sty1383}

\bibitem[{{Silich} {et~al.}(2007){Silich}, {Tenorio-Tagle}, \& {Mu{\~n}oz-Tu{\~n}{\'o}n}}]{Silich2007}
{Silich}, S., {Tenorio-Tagle}, G., \& {Mu{\~n}oz-Tu{\~n}{\'o}n}, C. 2007, \apj, 669, 952, \dodoi{10.1086/521706}

\bibitem[{{Silich} {et~al.}(2004){Silich}, {Tenorio-Tagle}, \& {Rodr{\'\i}guez-Gonz{\'a}lez}}]{Silich2004}
{Silich}, S., {Tenorio-Tagle}, G., \& {Rodr{\'\i}guez-Gonz{\'a}lez}, A. 2004, \apj, 610

\bibitem[{{Smith} \& {Wang}(2004)}]{Smith2004}
{Smith}, D.~A., \& {Wang}, Q.~D. 2004, \apj, 611, 881, \dodoi{10.1086/422181}

\bibitem[{{Smith} {et~al.}(2006){Smith}, {Westmoquette}, {Gallagher}, {O'Connell}, {Rosario}, \& {de Grijs}}]{Smith2006}
{Smith}, L.~J., {Westmoquette}, M.~S., {Gallagher}, J.~S., {et~al.} 2006, \mnras, 370, 513, \dodoi{10.1111/j.1365-2966.2006.10507.x}

\bibitem[{{Sukhbold} {et~al.}(2016){Sukhbold}, {Ertl}, {Woosley}, {Brown}, \& {Janka}}]{Sukhbold2016}
{Sukhbold}, T., {Ertl}, T., {Woosley}, S.~E., {Brown}, J.~M., \& {Janka}, H.~T. 2016, \apj, 821, 38, \dodoi{10.3847/0004-637X/821/1/38}

\bibitem[{{Takahashi}(2018)}]{Takahashi2018}
{Takahashi}, K. 2018, \apj, 863, 153, \dodoi{10.3847/1538-4357/aad2d2}

\bibitem[{{Tao} {et~al.}(2019){Tao}, {Feng}, {Zhang}, {Bu}, {Zhang}, {Qu}, \& {Zhang}}]{Tao2019}
{Tao}, L., {Feng}, H., {Zhang}, S., {et~al.} 2019, \apj, 873, 19, \dodoi{10.3847/1538-4357/ab0211}

\bibitem[{{Turner} {et~al.}(2015){Turner}, {Beck}, {Benford}, {Consiglio}, {Ho}, {Kov{\'a}cs}, {Meier}, \& {Zhao}}]{Turner2015}
{Turner}, J.~L., {Beck}, S.~C., {Benford}, D.~J., {et~al.} 2015, \nat, 519, 331, \dodoi{10.1038/nature14218}

\bibitem[{{Turner} {et~al.}(2017){Turner}, {Consiglio}, {Beck}, {Goss}, {Ho}, {Meier}, {Silich}, \& {Zhao}}]{Turner2017}
{Turner}, J.~L., {Consiglio}, S.~M., {Beck}, S.~C., {et~al.} 2017, \apj, 846, 73, \dodoi{10.3847/1538-4357/aa8669}

\bibitem[{{Vink}(2022)}]{Vink2022}
{Vink}, J.~S. 2022, \araa, 60, 203, \dodoi{10.1146/annurev-astro-052920-094949}

\bibitem[{{Vink} {et~al.}(2001){Vink}, {de Koter}, \& {Lamers}}]{Vink2001}
{Vink}, J.~S., {de Koter}, A., \& {Lamers}, H.~J.~G.~L.~M. 2001, \aap, 369, 574, \dodoi{10.1051/0004-6361:20010127}

\bibitem[{{Vink} \& {Sander}(2021)}]{Vink2021}
{Vink}, J.~S., \& {Sander}, A. A.~C. 2021, \mnras, 504, 2051, \dodoi{10.1093/mnras/stab902}

\bibitem[{{{\v{S}}urlan} {et~al.}(2013){{\v{S}}urlan}, {Hamann}, {Aret}, {Kub{\'a}t}, {Oskinova}, \& {Torres}}]{Surlan2013}
{{\v{S}}urlan}, B., {Hamann}, W.~R., {Aret}, A., {et~al.} 2013, \aap, 559, A130, \dodoi{10.1051/0004-6361/201322390}

\bibitem[{{Westmoquette} {et~al.}(2013){Westmoquette}, {James}, {Monreal-Ibero}, \& {Walsh}}]{Westmoquette2013}
{Westmoquette}, M.~S., {James}, B., {Monreal-Ibero}, A., \& {Walsh}, J.~R. 2013, \aap, 550, A88, \dodoi{10.1051/0004-6361/201220580}

\bibitem[{{Woosley} \& {Weaver}(1995)}]{Woosley1995}
{Woosley}, S.~E., \& {Weaver}, T.~A. 1995, \apjs, 101, 181, \dodoi{10.1086/192237}

\bibitem[{W\"unsch {et~al.}(2008)W\"unsch, Tenorio-Tagle, PalouÅ¡, \& Silich}]{Wuensch2008}
W\"unsch, R., Tenorio-Tagle, G., PalouÅ¡, J., \& Silich, S. 2008, \apj, 683, 683, \dodoi{10.1086/589967}

\bibitem[{{Zhang} {et~al.}(2008){Zhang}, {Woosley}, \& {Heger}}]{Zhang2008}
{Zhang}, W., {Woosley}, S.~E., \& {Heger}, A. 2008, \apj, 679, 639, \dodoi{10.1086/526404}

\end{thebibliography}
\bibliographystyle{aasjournal}

\end{document}